\documentclass[prl,twocolumn,superscriptaddress,showpacs,amssymb]{revtex4-1}

\usepackage{psfrag,graphicx}
\usepackage{tikz}
\usepackage{bbm}
\usepackage{pgfplots}
\usepackage{dcolumn}
\usepackage{bm}
\usepackage{amsfonts,amssymb,amsmath} 
\usepackage{xcolor}
\usepackage{tikz}
\usepackage{times}
\usetikzlibrary{decorations.markings}
\usetikzlibrary{arrows}
\usepackage{epstopdf}
\usepackage{soul}

\definecolor{jens}{rgb}{.2,0.7,.9}

\definecolor{konn}{rgb}{.4,0.2,.6}

\usepackage{mathbbol}

\newcommand{\be}{\begin{equation}}
\newcommand{\ee}{\end{equation}}
\newcommand{\bq}{\begin{eqnarray}}
\newcommand{\eq}{\end{eqnarray}}

\newcommand{\rf}[1]{(\ref{#1})}

\newcommand{\tr}{\mathrm{tr}}
\newcommand{\ket}[1]{\left |#1 \right\rangle}
\newcommand{\bra}[1]{\left \langle #1 \right |}

\newcommand{\R}{\mathbbm{R}}
\newcommand{\Z}{\mathbbm{Z}}
\newcommand{\id}{\mathbbm{1}}

\usetikzlibrary{positioning,shapes.misc}

\begin{document}

\title{Diagnosing Topological Edge States via Entanglement Monogamy}

\author{K.\ Meichanetzidis}\email{mmkm@leeds.ac.uk}
\affiliation{School of Physics and Astronomy, University of Leeds, Leeds LS2 9JT, United Kingdom}
\author{J.\ Eisert}
\affiliation{Dahlem Center for Complex Quantum Systems, Freie Universit{\"a}t Berlin, 14195 Berlin, Germany}
\author{M.\ Cirio}
\affiliation{  Interdisciplinary Theoretical Science Research Group (iTHES), RIKEN, Wako-shi, Saitama 351-0198, Japan }
\author{V.\ Lahtinen}
\affiliation{Dahlem Center for Complex Quantum Systems, Freie Universit{\"a}t Berlin, 14195 Berlin, Germany}		
\author{ J.\ K.\ Pachos}
\affiliation{School of Physics and Astronomy, University of Leeds, Leeds LS2 9JT, United Kingdom}

\date{\today}

\begin{abstract}
Topological phases of matter possess intricate correlation patterns typically probed by entanglement entropies or entanglement spectra. In this work, we propose an alternative approach to assessing topologically induced edge states in free and interacting fermionic systems. We do so by focussing on the fermionic covariance matrix. This matrix is often tractable either analytically or numerically and it precisely captures the relevant correlations of the system. By invoking the concept of monogamy of entanglement we show that highly entangled states supported across a system bi-partition are largely disentangled from the rest of the system, thus appearing usually as gapless edge states. We then define an entanglement qualifier that identifies the presence of topological edge states based purely on correlations present in the ground states.  We demonstrate the versatility of this qualifier by applying it to various free and interacting fermionic topological systems.
\end{abstract}

\maketitle

Consider a two-dimensional gapped system prepared in a pure state $\rho$ partitioned into region $A$ and its complement $B$. For large enough regions with smooth boundaries the entanglement entropy corresponding to the reduced density matrix $\rho_A=\tr_B (\rho)$ is expected to take the form $S(\rho_A) =(\alpha+\gamma)|\partial A| - \gamma + O(|\partial A|^{-\beta})$, where $\alpha,\beta, \gamma\geq 0$ are constants and $\partial A$ denotes the boundary of $A$~\cite{Area}. The first term describes the area law contribution that is generally considered to be non-universal since $\alpha$ depends on system specific microscopics and can change adiabatically~\cite{Budich14}. In contrast, the second term $\gamma$ is a universal constant called the topological entanglement entropy~\cite{Hamma,KitaevPreskill, LevinWen}. The numerical extraction of $\gamma$ has become a feasible numerical instrument to identify topologically ordered states in strongly correlated systems~\cite{Jiang12}. 

Topological phases of fermions, commonly referred to as topological insulators and superconductors, can have a band structure that is characterised by non-trivial topological indices~\cite{Schnyder08} even if $\gamma=0$. A physical consequence of this is the appearance of edge states at their boundaries~\cite{Qi}, that can be used as means to identify topological phases theoretically~\cite{Zhang09,Kitaev09} or in the laboratory~\cite{Koenig07}. Edge states are eigenstates of the Hamiltonian that are exponentially localised at the boundary of the system and minimally coupled to the rest of the system. Due to the equivalence of the entanglement spectrum \cite{Li08} --- the spectrum $\{ \epsilon_j \}$ of the virtual entanglement Hamiltonian $H_A $ defined by $\rho_A= e^{-H_A}/ \tr (e^{-H_A})$ --- with the physical energy spectrum \cite{Fidkowski10}, virtual entanglement edge states are also witnessed in the spectrum of $\rho_A$ in topological phases. The virtual edge states are also exponentially localised at the partition boundary $\partial A$, thus being minimally correlated to the bulk states of $A$ but highly entangled to the complementary subsystem $B$. While both the energy and entanglement spectra can be adiabatically tuned, topology implies that the virtual edge states persist unless the bulk energy gap closes. Hence, topological phases cannot be adiabatically connected to a product state with $\alpha =0$~\cite{Budich14}. However, to use this entropic criterion to identify topological phases requires full diagonalisation of the model, which is in general challenging, particularly in the presence of interactions.

Here we take an entirely different approach to identify edge states and topological phases in both free and interacting fermionic systems. In contrast to entropic witnesses that address collective mode effects, we focus on two-point correlations. These are conceptually simpler objects, and at the same time ones that can be numerically much more easily obtained by, e.g., tensor network \cite{Tensors}
or Monte Carlo methods \cite{Broecker,Grover13}. We show that the high entanglement of individual virtual edge states is efficiently captured by the fermionic covariance matrix even in interacting models. Our argument is based on the extremality property of Gaussian fermionic states conjunct with the notion of the monogamy of entanglement~\cite{Wootters,OsborneVerstraete}. In terms of entanglement entropy we show how to single out the area law contributions of the virtual edge states from two-point correlations, without the need to study system-size scaling or needing to adiabatically tune the Hamiltonian. Finally, we demonstrate that this correlation based signature, unlike spectral signatures of topological phases, is robust against perturbations, disorder and interactions.

{\bf \em The covariance matrix.} We first introduce the covariance matrix that facilitates our study of edge state correlations. A physical system that embodies $N$ fermionic modes with annihilation operators $f_1,\dots,f_N$ can always be associated with $2N$ Hermitian
Majorana fermions $\gamma_1,\dots, \gamma_{2N}$, by $f_j= {(\gamma_{2j-1}+i \gamma_{2j})}/{\sqrt{2}}$, for $j=1,\dots, N$. The second moments of such Majorana fermions of an arbitrary fermionic density matrix $\rho$ can be collected in the covariance matrix $\Gamma$ \cite{Fidkowski10,Peschel,Bravyi,Cramer}, 
whose elements are the two-point correlations
\be
\Gamma_{j,k} = i {\tr} (\rho{[}\gamma_j,\gamma_k{]}).  
\ee
This is a real $2N\times 2N$-matrix which is well defined for arbitrary fermionic states, including ground states of superconducting or interacting models, and allows to treat them all on an equal footing. Since it satisfies $\Gamma=-\Gamma^T$ and $\Gamma^T \Gamma \leq \id$, it has eigenvalues $\{\mu_j\}\in[-1,1]$ coming in positive and negative pairs. For an arbitrary bi-partition of the system in regions $A$ and $B$ the covariance matrix can be written as
\be
	\Gamma =  \left[
	\begin{array}{cc}
	\Gamma_A & \Gamma_{AB}\\
	-\Gamma_{AB}^T & \Gamma_B
	\end{array}
	\right] 
	,
\ee
with $\Gamma_{A(B)}$ reflecting the second moments of the reduced state $\rho_{A(B)}$ and $\Gamma_{AB}$ capturing correlations between $A$ and $B$. 

For free fermionic systems all ground states are Gaussian states and as such are completely defined by the covariance matrix 
$\Gamma$. The positive eigenvalues $\{ \mu_j^A \}$ of the covariance matrix $\Gamma_A$ are in one-to-one correspondence with the entanglement spectrum $\{ \epsilon_j \}$ through the relation $\mu^A_j=(1- e^{\epsilon_j})/(1+e^{\epsilon_j})$. For pure Gaussian states, one finds that the singular values $\xi_j^{AB}$ of $\Gamma_{AB}$ and the eigenvalues of $\Gamma_A$ satisfy $(\mu_j^{A})^2+ (\xi_j^{AB})^2=1$. The entanglement entropy can be evaluated by summing over the contributions from each mode
\be
S(\rho_A)=-{1 \over 2}
\sum_{j=1}^{2N}
 {1+\mu^A_j\over 2}\log{1+\mu^A_j\over 2}.
\label{SA1}
\ee
Thus the modes with $\mu^A_j=0$ (always coming in pairs)  are uncorrelated with the rest of $A$ and are maximally entangled with modes in $B$ witnessed by $\xi_j^{AB}= 1$. They correspond to the virtual edge states that translate to maximally entangled modes in the entanglement spectrum of topological free fermion systems, which contribute a maximal entropy of ${1 \over 2}\log (2)$ per mode~\cite{Fidkowski10}.  

We next turn to studying the properties of edge states of interacting fermions by considering the highly correlated modes with $\xi^{AB}_j\approx 1$. To relate these states to the virtual edge states we employ the monogamy of entanglement. This is a powerful general principle that allows us to detect edge states from the eigenvalues of the covariance matrix $\Gamma$. 

{\bf \em Entanglement monogamy.} Monogamy of entanglement states that no mode in $A$ maximally entangled with a mode in $B$ can be entangled with any other mode in $A$ or $B$~\cite{Wootters,OsborneVerstraete}. In the language of covariance matrices, singular values $\xi_j^{AB}=1$ of $\Gamma_{AB}$ imply an eigenvalue $\mu^A_j=0$ (note that the converse is not necessarily true). Thus such uncorrelated modes within $A$ must be decoupled from the bulk states and appear as exponentially localised states at the boundary of $A$.

To make this property more general and thus applicable to realistic systems we consider the concept of monogamy in the presence of high entanglement between $A$ and $B$ that is not necessarily maximal. We start by discussing a two-mode subsystem. Let $\Xi$ be any principal $4\times 4$-submatrix of the covariance matrix of an arbitrary bi-partite fermionic state with reduced state $\sigma$. This matrix can be brought into the form
\be
	\Xi =\left[
	\begin{array}{cccc}
	0 & a & 0 & b\\
	-a & 0 & c &0\\
	0 & -c & 0 &d\\
	-b & 0 & -d & 0\\
	\end{array}
	\right]  =
	\left[
	\begin{array}{cc}
	\Xi_A & \Xi_{AB}\\
	- \Xi_{AB}^T & \Xi_B
	\end{array}
	\right].
\label{eqn:cova}
\ee
This is a consequence of the real
special orthogonal singular value decomposition, applied to both local modes individually. 
This covariance matrix corresponds to a pure maximally entangled state exactly if $|b|=|c|=1$, as only then
$\Xi^T\Xi=\id$.
Invoking the Jordan-Wigner transformation, the state $\sigma$ can be written as 
$\sigma= (1-\epsilon) \omega+ \epsilon \eta$, where $\omega$ is a maximally entangled state of minimum dimension, $ \min(|b|,|c|) > 1-\epsilon \ $ is the smallest singular value of $\Xi_{AB}$, and $\eta$ an orthogonal residual state. In other words, we can argue about the weight of a maximally entangled state by considering only the covariance matrix rather than the full state $\sigma$ of the system. In particular, if $ \min(|b|,|c|)>1-\epsilon$, then
\be
	\| \sigma-\omega \|_1 \leq 2\epsilon,
\ee
with the trace distance defined as $\|A\|_1= \tr(|A|)$ for operators $A$. Hence, by considering the eigenvalues of $\Xi_{AB}$ we can deduce how close the state $\sigma$ of the system is to a maximally entangled state $\omega$. 
The same argument applies to an arbitrary number of modes. If $2k$ singular values of $\Xi_{AB}$ are larger
than $1-\epsilon$, then one can identify a subspace embodying $k$ pairs of fermionic modes that are in trace distance closer than $2\epsilon$ to $k$ maximally entangled pairs.

The almost maximally entangled modes are largely disentangled from the remaining fermionic modes, as 
dictated by the monogamy of entanglement~\cite{Wootters,OsborneVerstraete}. To make this notion more precise let us first focus on 
the situation of the reduced state
supported on modes $S_1$ and $S_2$
being in a close to maximally entangled state; this is meant in the sense that this reduced state
can be written as $\sigma=  (1-\epsilon) \omega+ \epsilon \eta$ as above.
Then the mode  $S_1$ will be little entangled with any other individual mode of the system. In fact,
the sum of all entanglements of formation $E_F(1:j)$ \cite{Bennett} between $S_1$ and any other mode  $S_j$ is upper bounded by
\be
	\sum_{j=2}^N 
	E_F(1:j)\leq \log(2)(1 -(1-\epsilon)^2 ),
\ee
and is hence small if $\epsilon$ is close to zero. This is a consequence of the following facts. The entanglement of formation $E_F$ and the tangle $\tau$ \cite{Wootters,OsborneVerstraete} are related as $E_F^2\geq \tau$. Then, the entanglement of formation is an entanglement monotone, so that $E_F(1:2)\geq 1-\epsilon$ holds true. Finally, we make use of the monogamy of entanglement inequality for the tangle as discussed in Ref.\ \cite{OsborneVerstraete}.

The close to maximally entangled pair on $S_1$ and $S_2$ is also monogamous and disentangled  in a different sense.
These pairs are as a whole minimally entangled with all other modes, in the sense that
\be
	E_F(1,2: 3,\dots, N)\leq 2 \log(2)\epsilon.
\ee
This bound is derived by using the convexity of the entanglement of formation and noting that the maximally entangled  (Dirac fermion) pair takes the value $\log (2)$ in the chosen convention.
Again, it is straightforward to generalise this argument to the case where
$2k$ singular values are larger than $1-\epsilon$. Then the respective $k$ modes are at most 
$2 k\log(2)\epsilon$ entangled with the modes forming the complement of the system.

This result is general and applies to both free and interacting fermionic systems alike. It states that, due to their maximal correlations across $\partial A$, virtual edge modes appear as largely disentangled from the rest of the system. In 1D systems this decoupling dictates that the edge states appear as zero modes in the entanglement spectrum. In 2D or 3D systems they appear as gapless states freely propagating at $\partial A$.

{\bf \em Entropic lower bound.} The existence of virtual edge states implies a lower bound for the entanglement entropy~\cite{Klich04}. In terms of the correlation part of the covariance matrix~\cite{supp}
\be
	S(\rho_A)\geq {1 \over 2}  \|\Gamma_{AB}\|_2^2 \log (2),
\label{eqn:ob1}
\ee	
where the $2$-norm is defined as $ \|A\|_2^2= \tr(A^2)$, the sum of the squared singular values. Note that \rf{eqn:ob1} is general, holding for free and interacting systems alike. For its interpretation note that the contributions from the bulk states to the entanglement entropy can be adiabatically removed. But the fundamental properties of virtual edge states are resilient against any adiabatic evolution of the corresponding physical Hamiltonian~\cite{Schnyder08}. Since topological phases are characterised by $\|\Gamma_{AB}\|_2^2\neq 0$, then the lower bound \rf{eqn:ob1} becomes non-trivial dictating that the area law coefficient $\alpha$ can never be made zero.

{\bf \em Entanglement qualifier.} We have shown above that nearly maximally entangled modes of the system are witnessed by singular values $\xi^{AB}_j \approx 1$ of $\Gamma_{AB}$. In analogy with the entanglement gap~\cite{Thomale} these are separated by a \emph{covariance gap} from the $\xi^{AB}_j$ corresponding to non-universal bulk states. This gives an efficient diagnostic tool to probe the topological character of the system. To count the number of such modes in a way robust to imperfections and finite system sizes, we define the entanglement qualifier $\mathcal{S}_q$ for some positive integer $q$ as
\be
\mathcal{S}_q = \textrm{Tr} \left( \Gamma_{AB}^\dagger  \Gamma_{AB}^{} \right)^q.
\label{eqn:qualifier}
\ee
In the limit $q \to \infty$ this quantity converges to $M$ --- the number of maximally entangled modes in units of Majoranas (a Dirac mode counts as two Majoranas) ---  and thus detects the existence of physical edge states from the ground state. In the presence of a covariance gap, $\mathcal{S}_q$ gives the degeneracy of these states for large but non-infinite values of $q$.
 
To demonstrate that the qualifier $\mathcal{S}_q$ identifies the presence of edge states and the size of their akin Hilbert space, we apply it first to Kitaev's honeycomb lattice model (2D topological superconductor) \cite{Kitaev06} and Haldane's model (2D Chern insulator) \cite{Haldane88}, in the presence and absence of disorder. In these free models $\Gamma$ contains all the information about the ground state, while for interacting systems this is not the case. Nevertheless, we show in the context of the 1D Su-Schrieffer-Heeger model (SSH) \cite{Su79} with interactions that even if the spectral properties fail to identify virtual edge states, due to monogamy the covariance matrix successfully identifies their presence even for strong interactions.

\begin{figure}[t]
\includegraphics[width=0.235\textwidth]{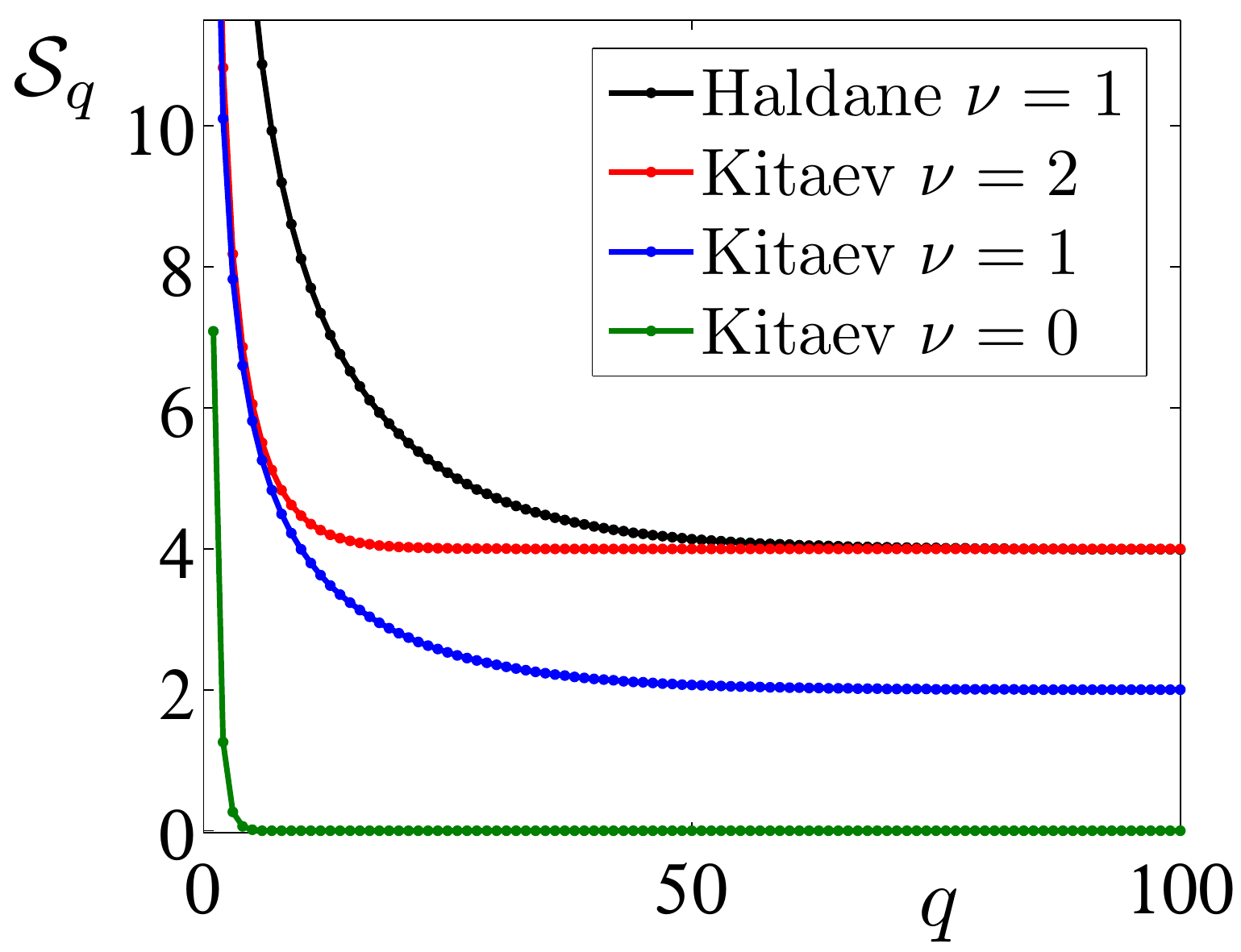}
\includegraphics[width=0.235\textwidth]{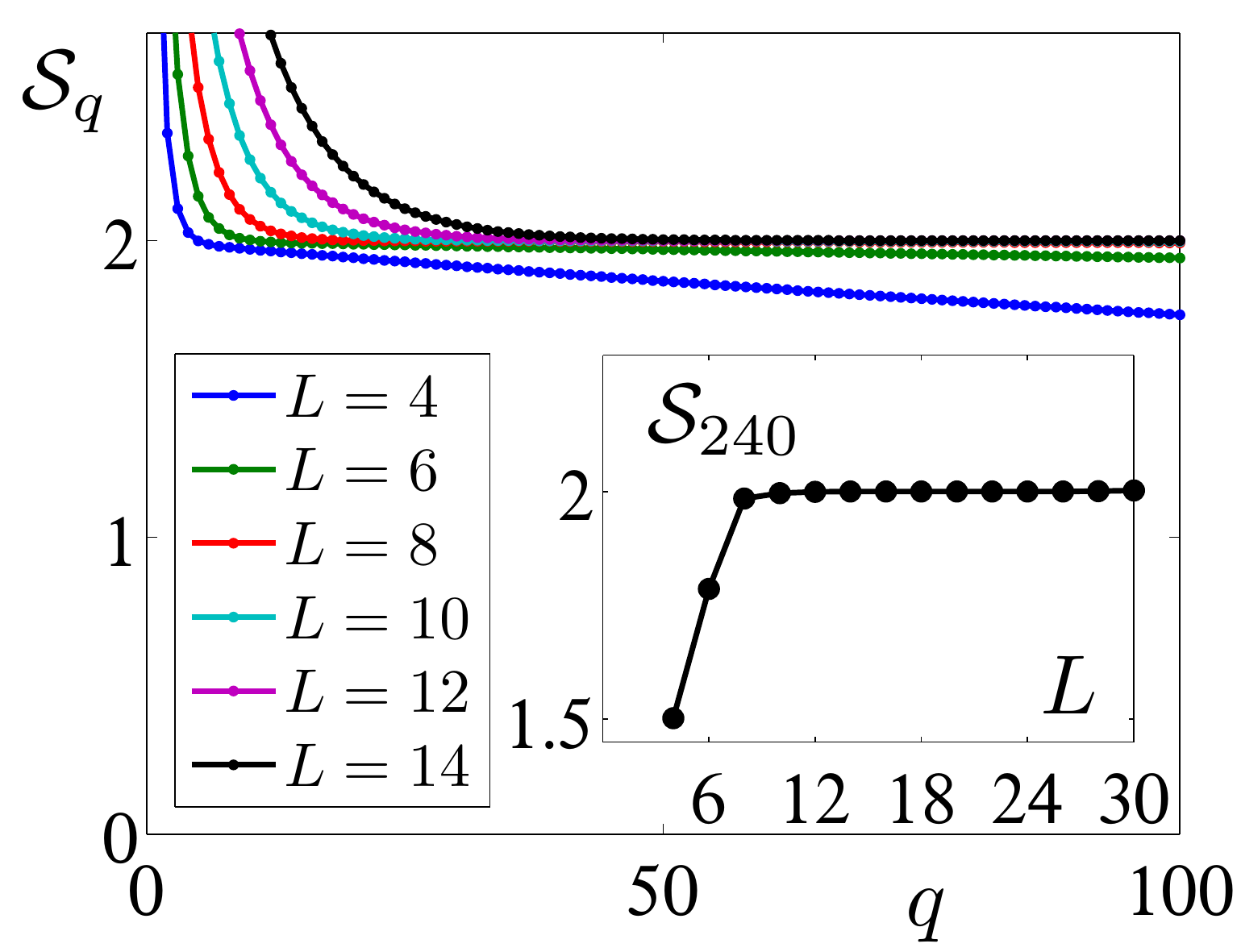}
\caption{\label{fig:converge} (Left) Convergence of qualifier ${\mathcal S}_{q}$ for the Kitaev and Haldane models with a two-component boundary $\partial A$. For Kitaev's model we find ${\mathcal S}_{q} \to 2\nu$ consistent with each phase supporting $\nu$ Majorana edge states per boundary. For the Haldane model we have ${\mathcal S}_{q} \to 4\nu$ as a Dirac mode corresponds to two Majorana modes. (Right) System size $L=L_x=L_y$ dependence of the qualifier ${\mathcal S}_{q}$ for Kitaev's model with $\nu=1$, in the presence of disorder of magnitude $\Delta=0.5$. The data is averaged over $50$ disorder realisations. The model parameters are given in Ref. \cite{supp}.}
\end{figure}

\begin{figure}[t] 
\begin{tabular}{cc}
\includegraphics[width=0.24\textwidth]{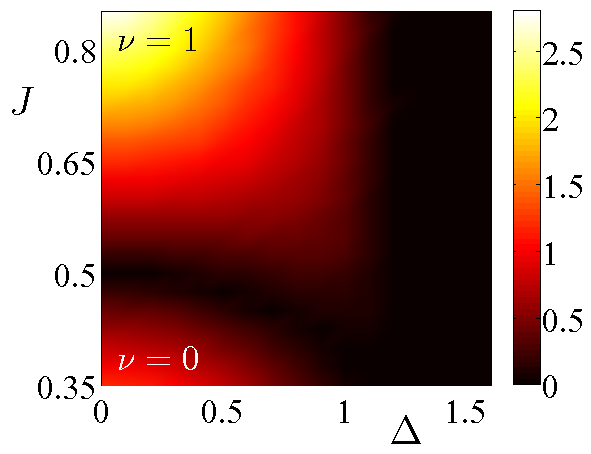} &
\includegraphics[width=0.24\textwidth]{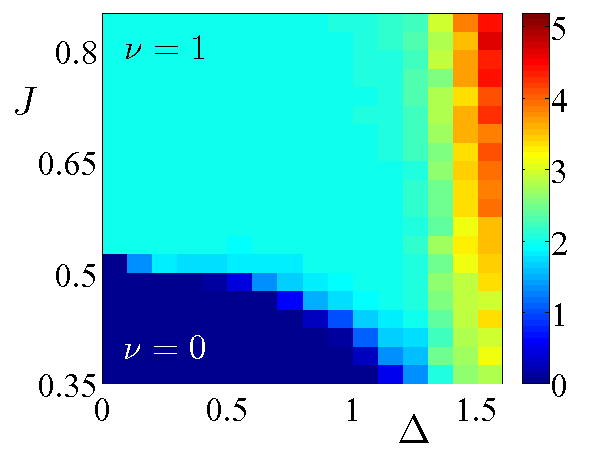} \\
Energy gap & Qualifier $\mathcal{S}_{240}$
\end{tabular}
\caption{\label{fig:PDs} (Left) Energy gap for the disordered vortex-free sector of Kitaev's model as a function of the coupling $J$ and the disorder $\Delta$. The topological phases with $\nu=0$ and $\nu=1$ are separated by a phase transition at $J_c=0.5$ for $\Delta=0$. The data is for $L_x=L_y=30$ averaged over $50$ disorder realisations of $J$ and $ K$~\cite{supp}. (Right) The same phase diagram diagnosed by $\mathcal{S}_q$ showing extended regions where $\mathcal{S}_{240}=2|\nu|$ identifies the topological phases even for strong disorder. The non-quantized behaviour for $\Delta > 1$ identifies the thermal metal phase~\cite{Lahtinen14}. }
\end{figure}

{\bf \em Free fermionic models.} 
Kitaev's model is equivalent to free Majorana fermions $\gamma_i$ on a honeycomb lattice coupled to a $\mathbb{Z}_2$ gauge field. The time-reversal symmetry broken variant is defined by the Hamiltonian \cite{supp}
\be \label{H_honey_Maj}
H = \frac{i}{2}\sum_{\langle i,j \rangle} J_{i,j} {u}_{i,j} \gamma_i \gamma_j + \frac{i}{2} K \sum_{\langle \langle i,j\rangle\rangle} {u}_{i,k} {u}_{k, j} \gamma_i \gamma_j,
\ee
where $J_{i,j}$ and $K$ are the nearest and next-nearest neighbour hopping amplitudes. The link variables $u_{i,j}=\pm 1$ are conserved quantities that define different vortex sectors that support various topological phases characterised by different Chern numbers $\nu$ \cite{Lahtinen12,Lahtinen14}. We consider three different regimes with distinct edge spectrum: (i) the $\nu=0$ Abelian phase with no edge states, (ii) the $\nu=1$ chiral non-Abelian phase with a single Majorana edge state, and (iii)~the $\nu=2$ chiral Abelian phase in the full-vortex sector with two Majorana edge states (see \cite{supp} for the parameter regimes). Despite their different Chern numbers, the topological entanglement entropy takes the same value of $\gamma=\log (2)$ for all of these cases and is thus unable to distinguish between them \cite{Yao10}. The full spectrum of eigenstates is readily obtained by exact diagonalisation. Constructing the full covariance matrix \cite{supp} and evaluating the qualifier $\mathcal{S}_{q}$ for each of the cases (i), (ii), and (iii) for a two component boundary $\partial A$, we find that it quickly converges to the values $M=0$, $M = 2$ and $M = 4$, respectively, as shown in Fig.~\ref{fig:converge}~(Left). This is in exact agreement with the Chern number of each phase~\cite{Lahtinen12}. Similar analysis is carried out for the Haldane model for complex fermions $f_i$,
\be
H=\sum_{\langle i,j \rangle} t_1 f^\dagger_i f_j + \sum_{\langle\langle i,j \rangle\rangle} t_2 e^{i \phi} f^\dagger_i f_j + \textrm{h.c.},
\label{eqn:HaldaneHam}
\ee
where $t_1,t_2>0$, and $\phi\in[-\pi,\pi]$. When this model is tuned to the Chern insulator phase characterized by $\nu=1$ \cite{supp}, Fig.~\ref{fig:converge} (Left) shows the qualifier $\mathcal{S}_q$ convering to $M=4$ consistent with the edge spectrum of a single Dirac fermion per edge. The qualifier $\mathcal{S}_q$ also detects topological edge states in the presence of local random disorder in the Hamiltonian couplings. Indeed, Fig.~\ref{fig:PDs}~(Right) shows that in the disordered Kitaev model the various topological phases are accurately identified with distinct quantised values of $\mathcal{S}_q$. We observe that disorder leaves the highly entangled states largely unaffected, mainly reducing slightly the singular values of $\Gamma_{AB}$ from the exact values $1$. Hence a large but finite value of $q$ is appropriate for robustly identifying the topological phases.

In the idealised case of exactly maximal entangled virtual edge modes one can take arbitrarily large values of $q$. Nevertheless, this is not always the case for finite-size systems or with the introduction of disorder, as shown Fig.\ \ref{fig:converge}~(right). In finite-size systems the edge states can hybridise leading to smaller entanglement between them, and hence a smaller lower bound. Moreover, the entanglement spectrum can exhibit even-odd effects in $|\partial A|$ that can wash out lower bound completely \cite{Wang15}. Nevertheless, due to the exponential localisation of the edge states, the lower bound can still be recovered in all cases via system-size scaling. Even-odd effects should vanish polynomially in $|\partial A|$, while decoupling of edge states occurs exponentially in the distance between boundary components. In the above examples optimal choices of parameters are to take even cut length and a square system. 

\begin{figure}[t]
\includegraphics[width=0.235\textwidth]{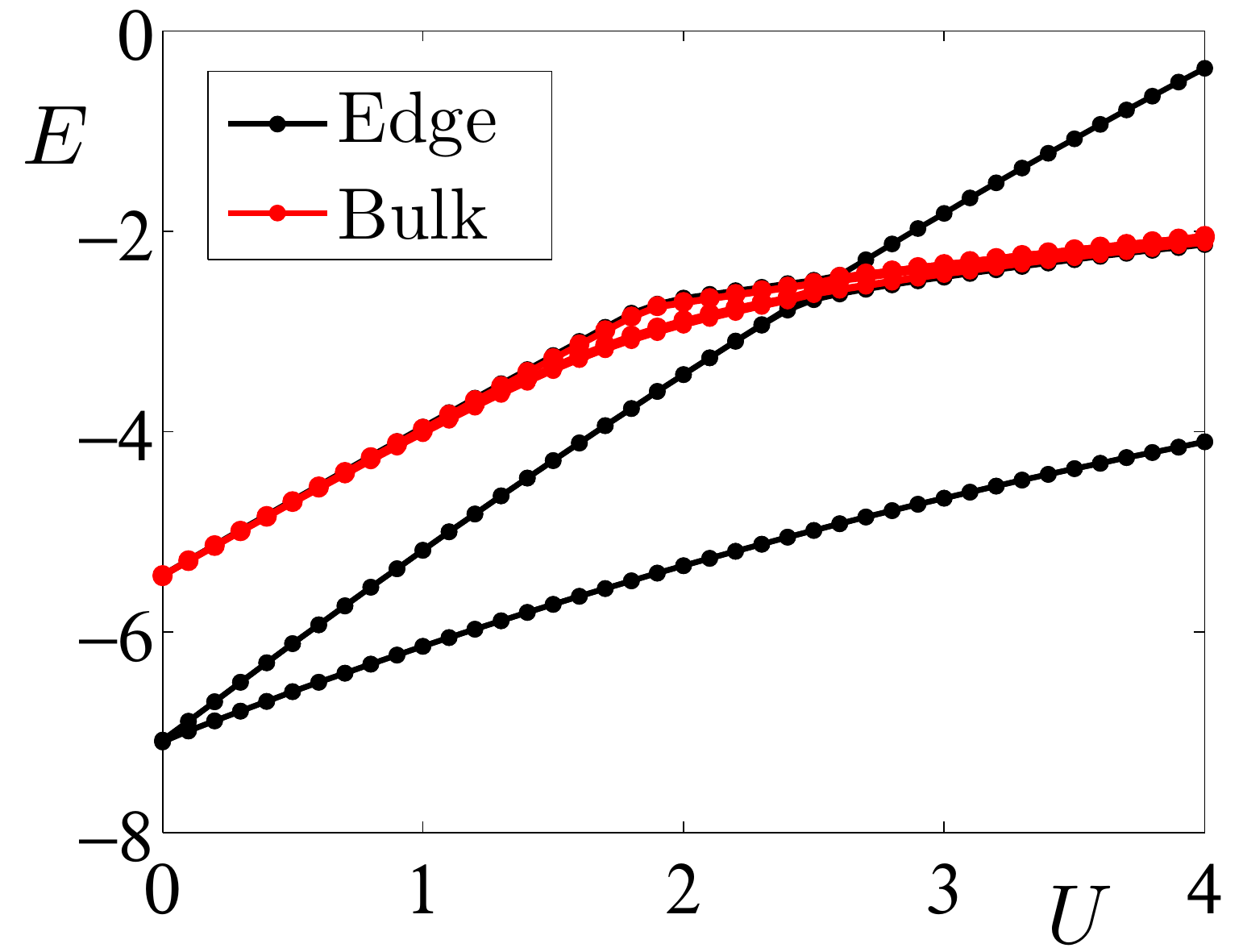}
\includegraphics[width=0.235\textwidth]{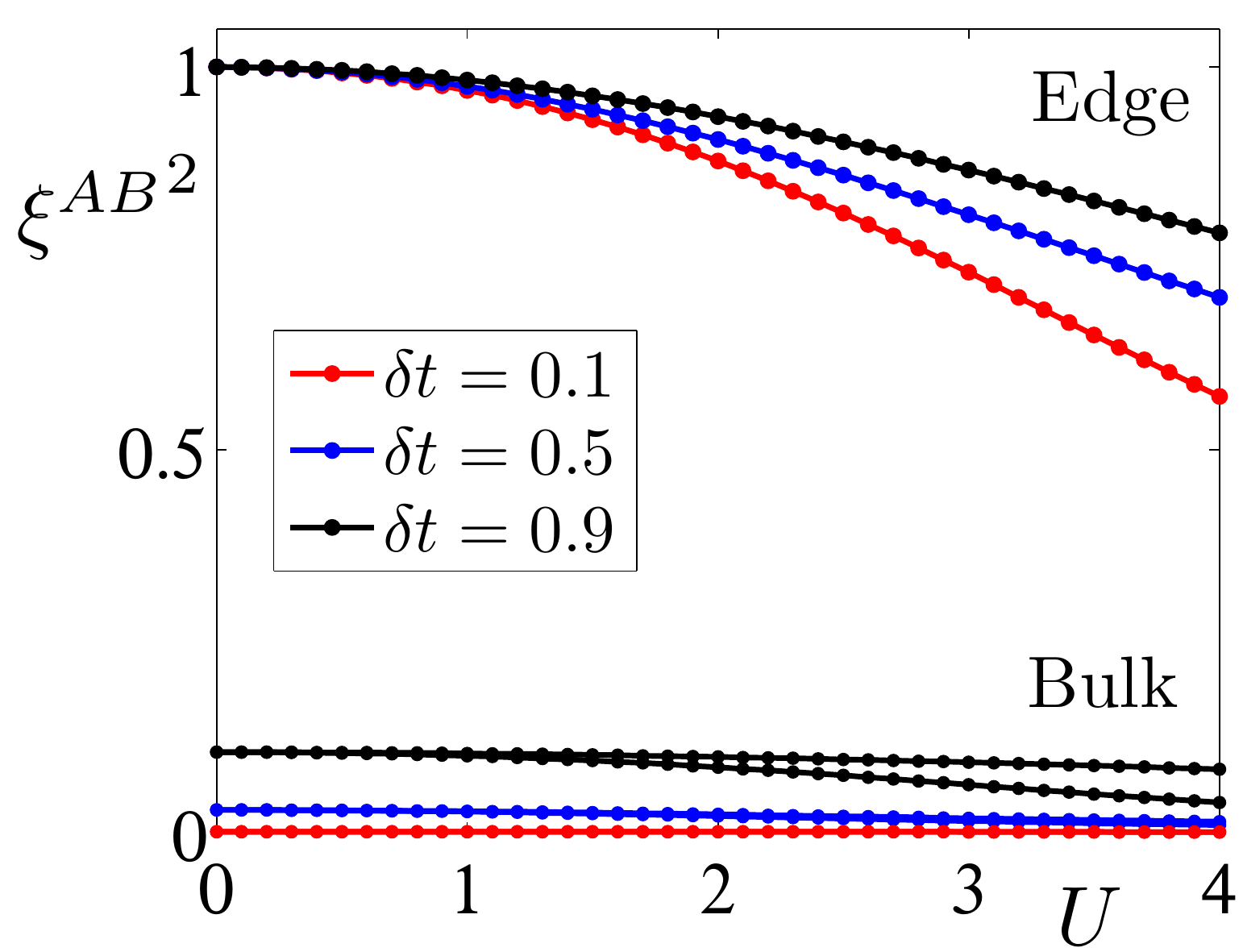}
\caption{\label{fig:interactions} (Left) Spectrum of the twelve lowest many-body states of the SSH model with interactions $U$ for $L=6$ and open boundaries terminating on weak bonds, $\delta t=0.75$. The two black lines are two-fold degenerate. The ground state degeneracy reduces from four-fold ($U=0$) to two-fold ($U>0$). (Right) The four-fold degenerate squared singular values ${\xi^{AB}_j}^2$ of $\Gamma_{AB}$ with periodic boundary conditions and equipartition (see Fig.\ 5 in Ref. \cite{supp}). While the degeneracy in the energy spectrum is lifted linearly with $U$ all the $M$ maximally entangled modes remain highly entangled and well separated from bulk entangled states by the covariance gap. Increasing $\delta t$ increases the entanglement of the edge states due to the decrease of the correlation length (equivalent to increasing the system size).}\end{figure}

{\bf \em Interacting fermion model.}
We turn now to the case of interacting fermions. We consider the 1D SSH model \cite{Su79} of spinful fermions on a chain of length $L$ with staggered hopping and on-site interactions of strength $U$,
\bq
H=&& \sum_{s} \sum^{L}_{i=1}-[t+\delta t(-1)^i] {f^{s}_{ i}}^\dagger {f^{s}_{ i+1}}+\text{h.c.}\\ \nonumber
&&+{U\over 2} \sum^L_{i=1} (n_{\uparrow , i} + n_{\downarrow  , i}-1)^2.
\eq
For periodic boundary conditions the ground state is unique, while for open boundary conditions and for $U=0$ edges terminating on weak bonds host an edge mode per spin component. We half-fill the system and restrict to the zero-spin sector which results in four-fold ground state degeneracy. Interactions linearly lift the degeneracy by hybridising the spinful edge states, resulting in a ground state with two-fold degeneracy~\cite{Wang15}, as shown in Fig.~\ref{fig:interactions}~(Left).

Let us consider the system in terms of correlations. When $U=0$, the model is one of free fermions and the covariance matrix $\Gamma$ contains all the information about the ground state. Partitioning a periodic system in half with a two-component boundary, the qualifier $\mathcal{S}_q$ converges for large $q$ to $M=4$, consistent with two Dirac fermion edge modes per strong bond cut by $\partial A$ \cite{supp}. For $U> 0$ the covariance matrix can still be readily evaluated from the eigenstates obtained via exact diagonalisation, though in contrast to the free case it does not contain all the information of the ground state. As shown in Fig.~\ref{fig:interactions} (Right), while the degeneracy of the energy is lifted, all four virtual Majorana modes that are highly entangled across $\partial A$ behave identically in the covariance spectrum, a behaviour not captured by the entanglement entropy~\cite{Wang15}. They remain at high entanglement and are separated from the bulk states by the covariance gap. This holds also in the extreme case, where $U$ is large enough to cause the energies of the edge states to cross the bulk energies. This behaviour of the qualifier $\mathcal{S}_q$ is consistent with the topological character of the system remaining unchanged. Indeed, there are no topological phase transitions and the winding number~\cite{DeLisle,Alba} remains the same when $U$ is introduced \cite{supp}. In other words, interactions only change adiabatically the edge spectrum and the edge modes remain well-defined. Thus the covariance matrix can faithfully detect edge states and thus identify topological phases also in the presence of interactions, where $\Gamma$ no longer fully characterises the ground state nor is in one-to-one correspondence with the entanglement spectrum. 

{\bf \em Conclusions.} Here we employed the covariance matrix to characterise topological phases in free and interacting fermion systems. We have shown that due to the monogamy of entanglement~\cite{Wootters,OsborneVerstraete} the number of highly entangled modes probes the number of topologically induced edge states, regardless of their energy. Hence, the covariance matrix, similar to the Green's function~\cite{Zhang09}, provides complementary information to the entanglement spectrum~\cite{Li08,Kitaev09}. We demonstrated that the high entanglement of the virtual edge states, unlike their spectrum, is a robust characteristic under perturbations that leave the topological phase unchanged, such as disorder or interactions. This gives a systematic and unambiguous way to study edge states and thus the topological character of fermionic systems in theoretical and numerical investigations as well as in experiments.

{\bf \em  Acknowledgements.} V.\ L.\ 
acknowledges the support by Dahlem Research School POINT Fellowship program,
K.\ M.\ by the EPSRC, J.\ E.\
by the ERC (TAQ), and the EU (AQUS, RAQUEL, SIQS).
We would like to thank
I.\ Klich,
G.\ Palumbo,
Z.\ Papic, and
C.\ Self
 for inspiring discussions.

\bibstyle{plain}

\newpage

\appendix
\section*{Supplementary material}

In this supplementary material we provide some additional detail to the way the covariance matrix is obtained for free-Majorana Hamiltonians, the proof of the entropic bound and
review the three models we discuss
in the main text: Kitaev's honeycomb model \cite{Kitaev06}, the Haldane model \cite{Haldane88}, 
and the SSH model \cite{Su79} with interactions.

\section{Covariance matrix for free Majorana fermions}
For any free and hence non-interacting Majorana Hamiltonian of $N$ modes
\be\label{hami}
H=i\sum_{i,j}A_{i,j}\gamma_i \gamma_j,
\ee
where $A= -A^T\in \R^{2N\times 2N}$ is antisymmetric, we find the covariance matrix of the ground state as follows~\cite{Kitaev06}.
We first construct the Majorana correlation matrix in the ground state
\be\label{eq:MajoranaCorrelations}
\langle \gamma_i \gamma_j \rangle=\delta_{i,j}-i B_{i,j},
\ee
where $B=Q \tilde{B} Q^T$. The matrix $Q$ is constructed in a way such that its columns are composed alternatively of the real and imaginary parts of the eigenvectors of $H$ and 
\begin{equation}
\tilde{B}=\bigoplus_{j=1}^N
	\left[
	\begin{array}{cc}
	0 & -1\\
	1 & 0
	\end{array}
	\right]. 
\end{equation}	
	Now from Eq.\ \eqref{hami} we obtain the correlation matrix $\Gamma_{i,j}=i \left(\langle \gamma_i \gamma_j \rangle-\langle \gamma_j \gamma_i \rangle\right)$
	for $i,j=1,\dots, 2N$.

\section{Entropy lower bound}
 In order to show the validity of (7) it is instructive to first consider a two-mode problem, with 
 quantum state $\sigma$. The covariance matrix of $\sigma$ has again the form
 \begin{equation}
	\Xi =\left[
	\begin{array}{cccc}
	0 & a & 0 & b\\
	-a & 0 & c &0\\
	0 & -c & 0 &d\\
	-b & 0 & -d & 0\\
	\end{array}
	\right]  =
	\left[
	\begin{array}{cc}
	\Xi_A & \Xi_{AB}\\
	- \Xi_{AB}^T & \Xi_B
	\end{array}
	\right].
\label{eqn:cova2}
\end{equation}
On using the Jordan Wigner transformation and the concavity of the von-Neumann entropy, one finds that 
\begin{equation}
	S(\sigma_A) \geq  \frac{1}{2}(|c+b|+ |c-b|). 
\end{equation}
The right hand side, in turn, can be lower bounded by 
$ \|\Xi_{AB}\|_2^2/2$, as an elementary argument shows. The generalisation of this to a bi-partitioned $N$ mode fermionic system 
with even $N$
can be performed by making use of the 
fact that by means of suitable special orthogonal mode transformations $O_A,O_B,U_A,U_B\in SO(N)$ 
local to $A$ and $B$, one can bring the off-diagonal block $\Gamma_{AB}$ into the form
\begin{equation}
	(O_A\oplus O_B) \Gamma_{AB} (U_A\oplus U_B)^\dagger = \bigoplus_{j=1}^{N/2} 
	\left[
	\begin{array}{cc}
	0 & b_j\\
	c_j & 0
	\end{array}
	\right].
\end{equation}	
Therefore, 
\begin{equation}
	S(\rho_A)\geq {1 \over 2}  \|\Gamma_{AB}\|_2^2 \log (2),
\end{equation}	
as stated in the main text. A related lower bound to the entropy in terms of Bogoliubov quasiparticles has been presented in Ref.\ \cite{Klich04}.

\section{Kitaev's honeycomb model} 

Kitaev's honeycomb model is a lattice model of spin $1/2$-particles residing on the vertices of a honeycomb lattice \cite{Kitaev06}. The spins interact according to the Hamiltonian
\be \label{H_honey}
	H = \sum_{\alpha =x,y,z} \sum_{\langle i,j \rangle} J_\alpha \sigma_i^\alpha \sigma_j^\alpha + K \sum_{\langle i,j,k \rangle} \sigma_i^\alpha \sigma_j^\beta \sigma_k^\gamma,
\ee
where $J_\alpha>0$ are nearest neighbour spin exchange couplings along links of type $\alpha$ and $K$ is the magnitude of a three spin term that explicitly breaks time reversal symmetry. The latter is required for the model to support gapped topological phases characterised by non-zero Chern numbers. For every hexagonal plaquette $p$ one can associate a $\Z_2$ valued six spin operator $\hat{W}_p=\sigma_1^x \sigma_2^y \sigma_3^z \sigma_4^x \sigma_5^y \sigma_6^z$ that describes a local symmetry $[H,\hat{W}_p]=0$. The Hilbert space of the spin model thus breaks into sectors labeled by the patterns $W=\{ W_p \}$ of the eigenvalues of $\hat{W}_p$. We refer to these sectors as {\it vortex sectors}, since $W_p=-1$ corresponds to having a $\pi$-flux vortex on plaquette $p$.

The interacting spin system \rf{H_honey} can be mapped to a system of Majorana fermions $\gamma_i=\gamma_i^\dagger$ 
coupled to a $\Z_2$ gauge field $\hat{u}_{i,j}$. The corresponding Hamiltonian is then given by
\be \label{H_honey_Maj}
	H = \frac{i}{2}\sum_{\langle i,j \rangle\in\Lambda} J_{i,j} {u}_{i,j} \gamma_i \gamma_j + \frac{i}{2} K \sum_{\langle \langle i,j\rangle\rangle\in\Lambda} {u}_{i,k} {u}_{k, j} \gamma_i \gamma_j,
\ee
where the first sum is over nearest neighbour sites $\langle i,j \rangle$, $J_{i,j}$ equals $J_x$, $J_y$ or $J_z$ depending on the type of link, and the second over next nearest neighbors $\langle\langle i,j \rangle\rangle$ with $k$ denoting the connecting site. The gauge field is static, i.e., the local gauge potentials satisfy $[H,\hat{u}_{i,j}]=0$. The plaquette operators become $\Z_2$ valued Wilson loop operators 
\begin{equation}
	\hat{W}_p=\prod_{(i,j)\in p} \hat{u}_{i,j}, 
\end{equation}	
which justifies the interpretation of the eigenvalues $W_p=-1$ corresponding to the presence of a $\pi$-flux vortex on plaquette $p$. By choosing a gauge, i.e., replacing the operators $\hat{u}_{i,j}$ with their eigenvalues $u_{i,j}=\pm1$, one restricts to a particular vortex sector $W(u)$. In each sector the Hamiltonian $H_{W(u)}$ is quadratic in the Majorana fermions and hence readily diagonalisable, with the resulting spectrum of free fermions depending only on the vortex sector $W$.

Depending on the couplings $J_{i,j}$ and $K$, and on the vortex sector $W$, the model supports several topological phases characterised by different Chern numbers $\nu$. In the vortex-free sector where $W_p=1$ on all plaquettes, and when $J_z > J_x+J_y$, or permutations thereof, the system is in a gapped Abelian phase with $\nu=0$. On the other hand, for isotropic couplings $J_x \approx J_y \approx J_z$ the system is in a gapped non-Abelian phase with $\nu=$sign$(K)$. In this phase vortices bind localised Majorana modes and behave thus as non-Abelian anyons. In the full-vortex sector ($W_p=-1$ on all plaquettes), the non-Abelian phase appearing for isotropic couplings is replaced by a chiral Abelian phase with $\nu=$sign$(K) 2$.
The full-vortex sector is implemented by fixing the gauge such that the $\mathbb{Z}_2$ filed value alternates on links of type $z$ along a row of the honeycomb lattice. Sparse vortex lattices also support various other phases whose nature depends on the vortex lattice spacing \cite{Lahtinen12, Lahtinen14}. Each topological phase of the honeycomb models supports $|\nu|$ Majorana edge states.
The parameters used to produce the data shown in Fig.\ 1 are $J=1$ for non-abelian vortex-free and $J=0.35$ for toric code phases, $K=0.15$, and $L_x=L_y=16$.

Disorder is introduced to the model by making the couplings $J$ and $K$ local random variables. In particular, we set ${J^{\alpha}_\text{d}}_{i,j} = {J^{\alpha}}_{i,j}(1+{\delta J}_{i,j})$, where ${\delta J}_{i,j}$ is drawn from the uniform centred box distribution $|{\delta J}_{i,j}|\leq \Delta$ of width $\Delta\geq 0$. The couplings $K$ are randomised similarly as ${K_\text{d}}_{i,j}=K(1+{\delta K}_{i,j})$, where ${\delta K}_{i,j}$ with $|{\delta K}_{i,j}|\leq {\Delta}^2$. The data for the disordered $\nu=1$ non-abelian phase presented in Fig.\ 1 (Right) was produced for $J=0.8$, $K=0.15$.


\section{Haldane model}

The Haldane model ~\cite{Haldane88} is also defined on a honeycomb lattice and describes the free fermions with broken time-reversal symmetry due to complex next-nearest neighbour hopping that does not introduce net flux to the system. The Hamiltonian is given by
\be
H=\sum_{\langle i,j \rangle \in \Lambda} t_1 f^\dagger_i f_j + \sum_{\langle\langle i,j \rangle\rangle \in \Lambda} t_2 e^{i \phi} f^\dagger_i f_j + \textrm{h.c.},
\label{eqn:HaldaneHam}
\ee
where $t_1$ and $t_2$ are real nearest- and next-nearest neighbour tunnelling amplitudes. The sign of $\phi\in [0,2\pi)$ is consistent with a chosen chirality. When $\phi \neq 0,\pi$ the system is in a gapped topological phase characterised by Chern number $\nu=\pm 1$ and exhibits a single Dirac edge state per boundary. The parameters used to produce the data shown in Fig.\ 1 (Left) are $t_1=1$, $t_2=1/3$, $\phi=\pi/3$, and $L_x=L_y=12$.

\section{SSHH model}
The  Hamiltonian describes $N_\uparrow$ and $N_\downarrow$ spinful fermions hopping on a chain of $L$ sites with alternating weak-strong hopping ~\cite{Su79} and interacting on site with a Hubbard interaction between spin components,
\bq
H=&& \sum_{s} \sum^{L}_{i=1}-(t+\delta t(-1)^i) {f^{s}_{ i}}^\dagger {f^{s}_{ i+1}}+\text{h.c.}\\ \nonumber
&&+U/2 \sum^L_{i=1} (n_{\uparrow, i} + n_{\downarrow, i}-1)^2,
\eq
where  $s=\uparrow, \downarrow$ and $n^s_i={f^{s}_{ i}}^\dagger {f^{s}_{ i}}$ populations for each spin with $N_s=\sum_i^L n^s_i$.
The total spin is defined as $S_z= (N_\uparrow-N_\downarrow)/2$. The Hamiltonian conserves the particle number and the total spin. Thus it splits into sectors labeled by $S_z$ determined by numbers of particles which can take the values $N_\uparrow=0,1,\dots,L$ and $N_\downarrow=L-N_\uparrow$. 

When $U=0$ the model is two copies of the SSH model of polyacetiline, one copy for each spin component.
There are $L-2$ states in the lower band and $L-2$ in the upper band and four zero-energy modes.
The lower band is fully occupied and the many-body ground state degeneracy is the number of ways $2$ particles can occupy four zero-modes, which is six. Two of these many-body states in the ground space have $S_z= \pm1$ and the other four have $S_z=0$.
Restricting to the $S_z=0$ sector, we are left only with four states in the ground space.

\begin{figure}[t]
\includegraphics[width=0.47\textwidth]{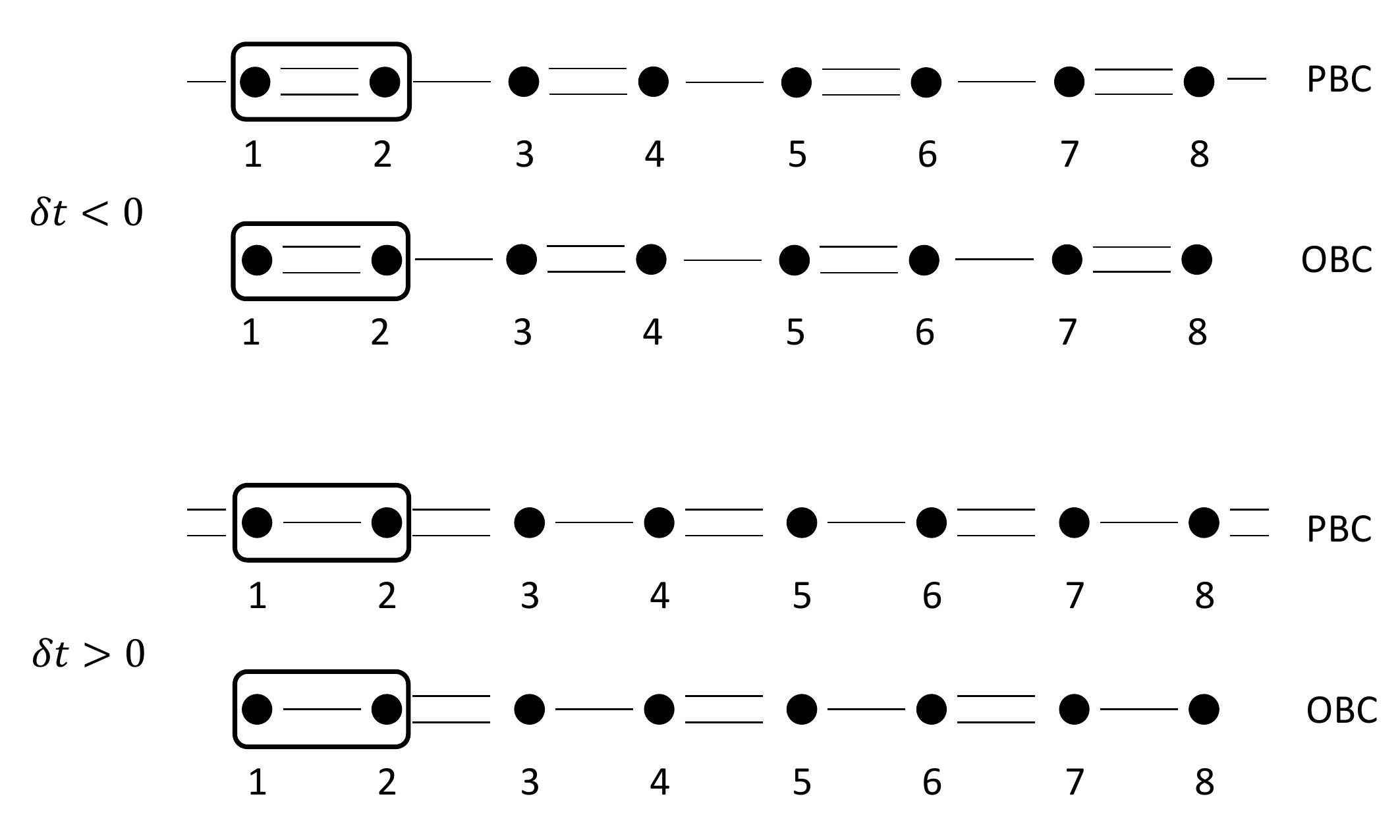}
\caption{ Representation of the alternating weak (single line) - strong (double line) hoppings of the SSHH model for open and periodic boundary conditions (OBC and PBC). The unit cell is depicted by a box containing two sites, one with an odd and one with an even index. The sign of $\delta t$ determines whether the unit cell contains a weak or a strong hopping, $\delta t>0$ ($\delta t<0$) for topological (trivial). For $\delta t>0 $ and OBC we have edge states exponentially localised on sites $1$ and $8$ where the chain terminates on weak bonds.}
\end{figure}


\subsection{Covariance matrix for the SSHH model}
We work at half filling so we choose $L$ to be even and $N_\uparrow+N_\downarrow=L$ and enumerate the unit cells with $x=1,\dots,L/2$. 
The Fock basis is $\{|b^\uparrow_1,  \dots ,b^\uparrow_L ,b^\downarrow_1 ,\dots ,b^\downarrow_L\rangle\}$ where $b^s_i\in\{0,1\}$, or fermionic occupations on each site, for spin $s=\uparrow,\downarrow$. The constraint for half filling and $S_z=0$ demands $\sum_{i , s} b^s_i=L$ and $\sum_{i} b^\uparrow_i=\sum_{i} b^\downarrow_i$. The dimension of the Fock space is the number of the binary words
that satisfy these two conditions.

After exact diagonalisation for periodic boundary conditions, we find the unique ground state vector to be a superposition on this basis
\begin{equation}
\ket{gs_{PBC}}=\sum_{\{ b^\uparrow , b^\downarrow \}}  A_{\{ b^\uparrow , b^\downarrow \} } \ket{\{ b^\uparrow , b^\downarrow \}}
\end{equation}
upon which the fermionic operators $f^{s}_i$ act.
From the correlation matrices $C^{s , s'}_{i,j}=\langle{f^s_i}^\dagger f^{s'}_j  \rangle$ and $\tilde{C}=\langle f^s_i f^{s'}_j \rangle$,
which are expressed in terms of the amplitudes $A_{\{ b^\uparrow , b^\downarrow \} }$
we get $\langle \gamma_k \gamma_l \rangle$ and so we compute $\Gamma$.

The data for Fig.\ 3 were produced for $L=6$ and the entanglement cut was performed as shown in Fig.\ 5. The number $M$ of singular values $\xi^{AB}=1$ depend on how we cut. In particular, $M=4  b_s $ maximally entangled Majorana pairs, where we have a factor of $2$ due to one Dirac fermion corresponding to two Majorana fermions, another factor of $2$ due to the two spin components, and $b_s$ is the number of strong bonds (double lines in Fig.\ 5) present at the boundary. In the $L=6$ equi-partite case we consider in the text we have $b_s=1$.

\begin{figure}[t]
\includegraphics[width=0.38\textwidth]{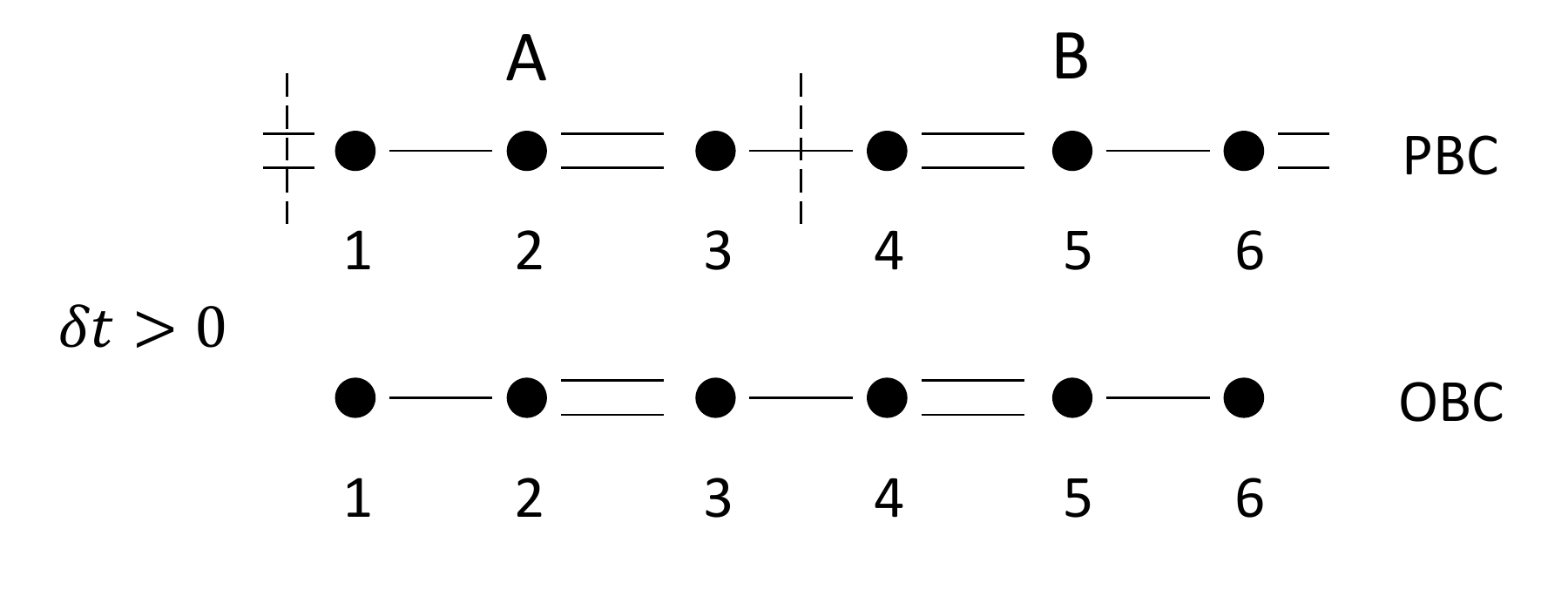}
\caption{ Representation of the SSHH alternating hoppings for $L=6$ with $\delta t>0$. For open boundary conditions (OBC) we have edge states and a four-fold ground state degeneracy due to zero-energy modes per spin component per edge. For periodic boundary conditions (PBC) we perform the entanglement cut at the positions indicated by the dashed lines.}
\end{figure}

\subsection{Winding number for the SSHH model}
SSHH supports a non-trivial winding number $w=\pm 1$ when $\delta t>0$.
We want the winding number as the winding of a unit vector $\langle {\bf\Sigma}^s(p)\rangle$ on the surface of the sphere, where $p\in[0,2\pi)$ is the momentum in the Brilluin zone.
In Fig.\ 4 we depict the SSHH hopping structure. We rename the fermionic operators that act on either site in the unit cell as $f^{s}_{2x-1}\rightarrow a^{s}_x, f^{s}_{2x}\rightarrow b^{s}_x$.
We define the observables that will give us a winding number for each spin component as ~\cite{DeLisle,Alba}
$ \Sigma^s_{x}={a^s}^\dagger_p {b^s}_p+{b^s}^\dagger_p {a^s}_p $, $ \Sigma^s_{y}=-i {a^s}^\dagger_p {b^s}_p +i {b^s}^\dagger_p {a^s}_p $, $ \Sigma^s_{z}= {a^s}^\dagger_p {b^s}_p - {b^s}^\dagger_p {a^s}_p $, where $s=\uparrow,\downarrow$.
In order to calculate $\bra{gs_{PBC}} {\bf \Sigma}^s (p) \ket{gs_{PBC}}$, we Fourier transform the fermionic operators ${a^s}_p, {b^s}_p$  back to real space
to obtain
\begin{align}
\Sigma^s_{x}&=\sum_{x',x=1}^{L/2}  C^{s,s}_{2x'-1,2x} e^{i p (x-x')}+C^{s,s}_{2x,2x'-1}  e^{-i p (x-x')} ,\\
\Sigma^s_{y}&=\sum_{x',x=1}^{L/2} -i C^{s,s}_{2x'-1,2x} e^{i p (x-x')} +i C^{s,s}_{2x,2x'-1} e^{-i p (x-x')},\\
\Sigma^s_{z}&=\sum_{x',x=1}^{L/2} C^{s,s}_{2x'-1,2x'-1} - C^{s,s}_{2x,2x} .
\end{align}
Each vector $\langle {\bf\Sigma}^\uparrow (p)\rangle$, $\langle {\bf\Sigma}^\downarrow (p)\rangle$ gives a winding number $w=1$ ($w=0$) when $\delta t>0$ ($\delta t<0$) for both $U=0$ and $U>0$. In Fig.\ 6 we show how the winding number is robust under interactions in the topological regime ($\delta t>0$).

\begin{figure}[t]
\includegraphics[width=0.3\textwidth]{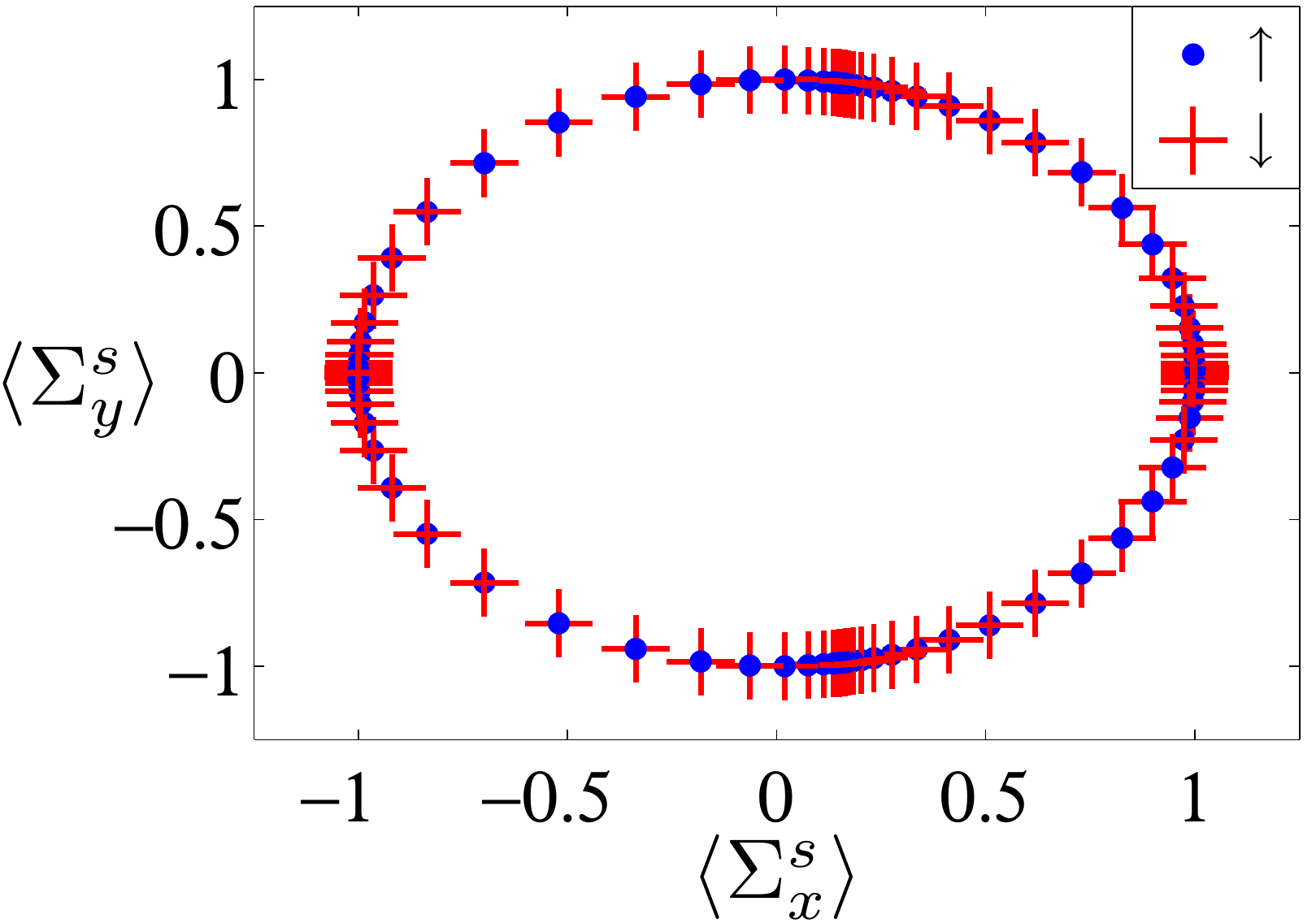}
\caption{Winding of the vectors $\langle {\bf \Sigma}^\uparrow \rangle$ (blue dot) and $\langle {\bf \Sigma}^\downarrow \rangle$ (red cross) as the momentum $p$ is varied from $0$ to $2\pi$ for the interacting SSH model for $L=8$ and $U=2$. The vectors wind on the $x-y$ plane. The winding that is present in the non-interacting case $U=0$ persists when interactions are turned on $U=2$. }
\end{figure}

\end{document}